\documentclass[twocolumn,floatfix,superscriptaddress,nofootinbib,amssymb,amsmath,aps,prc]{revtex4-1}

\usepackage[dvips]{graphicx}
\usepackage{latexsym,epsfig,bm,psfrag,subfigure}
\usepackage{color}         
\usepackage[bookmarksnumbered,bookmarksopen,colorlinks,citecolor=blue,linkcolor=blue]{hyperref} %

\newcommand{\eq}[1]{Eq.~(\ref{#1})}

\newcommand{\tpm}{\theta_{\ms +-}}
\renewcommand{\vec}{\boldsymbol}
\newcommand{\usl}[1]{U_{\ms 1#1}}

\newcommand{\mr}{M}

\newcommand{\ep}{$e^+$-$e^-$}
\newcommand{\lis}{{}^{7}\mathrm{Li}}
\newcommand{\bee}{{}^{8}\mathrm{Be}}

\newcommand{\ms}{\scriptscriptstyle}

\newcommand{\Mone}{\vec{\mathcal{O}}_{\mathrm{M}1}}
\newcommand{\Eone}{\vec{\mathcal{O}}_{\mathrm{E}1}}
\def\bea{\begin{eqnarray}}
\def\eea{\end{eqnarray}}
\def\g{\gamma}\def\m{\mu}\def\t{\tau}\def\ve{\varepsilon}\def\l{\lambda}\def\s{\sigma}\def\n{\nu}\def\o{\omega}

\begin{document}



\title{Can a protophobic vector boson explain the ATOMKI anomaly? }

\author{Xilin Zhang}
\email{zhang.10038@osu.edu}
\affiliation{Department of Physics, The Ohio State University, Ohio 43210, USA} 

\author{Gerald A. Miller}
\email{miller@phys.washington.edu}
\affiliation{Department of Physics, University of Washington, Seattle, WA \ \ 98195, USA}

\date{\today}

\begin{abstract}
In 2016, the ATOMKI collaboration announced  [PRL {\bf 116}, 042501 (2016)] observing  an unexpected enhancement of the \ep pair production signal in one of the $\bee$ nuclear transitions induced by an incident proton beam on a $^7$Li target. Many beyond-standard-model physics explanations have subsequently  been  proposed. One  popular theory is that the anomaly is caused by the creation of a protophobic vector boson ($X$) with  a mass around 17 MeV  [e.g., PRL\ {\bf 117}, 071803 (2016)] in the nuclear transition. We study this hypothesis by deriving  an isospin relation between photon and $X$ couplings to nucleons. This  allows us to find simple relations between protophobic $X$-production cross sections and those for  measured photon production. The net result is  that  $X$ production is  dominated by direct transitions induced by $E1^X$ and $L1^X$ (transverse and longitudinal electric dipoles) and $C1^X$ (charge dipole) without going through any nuclear resonance ({\it i.e.} Bremsstrahlung radiation) with a smooth  energy dependence that occurs for all proton beam energies above threshold.  This contradicts the experimental observations and invalidates  the  protophobic vector boson explanation.
\end{abstract}



\maketitle

\begin{figure}[h]
	\centering
	 \includegraphics[width=0.3 \textwidth]{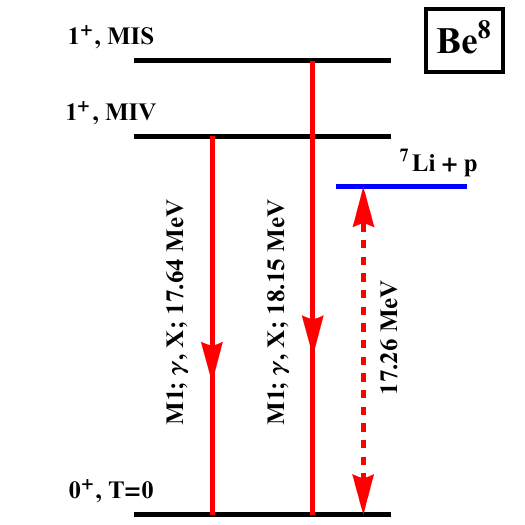}
     \caption{The $\bee$ levels~\cite{Tilley:2004zz} that are relevant for the M1 transitions producing photon ($\g$) and recently proposed vector boson $X$~\cite{Feng:2016jff,Feng:2016ysn,Feng:2020mbt}. The two $1^+$ resonance states are either mostly isovector (MIV) or mostly isoscalar (MIS). The blue line is the $\lis + p$ threshold. Note $X$ and (off-shell) $\g$ can further decay into \ep.} \label{fig:leveldiag}
\end{figure}

Ref.~\cite{Krasznahorkay:2015iga} observed an anomaly in  measuring  \ep pair production in  $\bee$'s nuclear transition between the 18.15 MeV  $1^+$ resonance and its $0^+$ ground state.  Fig.~\ref{fig:leveldiag} shows the relevant energy  levels~\cite{Tilley:2004zz}.  The two $1^+$ resonances are barely above the $\lis + p$ threshold. 
 The unexpected enhancement of the signal was observed in the large \ep invariant mass region (about 17 MeV) and in the large pair-correlation angles (near $140^\circ$) region.
The large angle enhancement is a simple kinematic signature of the decay of a heavy particle into an $e^+-e^-$ pair.
 The anomaly  has generated many beyond-standard-model physics explanations 
 ({\it e.g.},~\cite{Krasznahorkay:2015iga,Feng:2016jff, others}). 

Our focus is on the protophobic vector boson explanation (see {\it e.g.}~\cite{Feng:2016jff,Feng:2016ysn,Feng:2020mbt}). 
We shall show that taking this hypothesis seriously leads to the result that the large angle enhancement of pair-production would have been seen at all ATOMKI energies above threshold.

The physics of a boson that almost does not interact with protons provides an interesting contrast with  photon-nucleon interactions. We next show that isospin symmetry enables the derivation of a useful relation between the matrix elements of the two interactions.

 The photon-quark interactions are given by the following electromagnetic (EM) current in its $2^\mathrm{nd}$ quantization form: 
\begin{align}
j^\mu_\g & =\overline{Q}\g^\m\left({1\over 6}+{\t_3\over 2}\right)Q  = j^\mu_s + j^\mu_v \ ,\ \mathrm{with} \notag \\  
 j^\mu_s & \equiv \overline{Q} \g^\m { 1 \over 6}Q \ ,\ j^\mu_v  \equiv \overline{Q}\g^\m{\t_3\over 2}Q   \  .
 \label{g}
\end{align}
Here, $Q\equiv (u\ ,d)^T$ is the iso-doublet quark field operator $\t_3$ the isospin Pauli matrix, and $j^\mu_v$ and $j^\mu_s$ as the isovector and isoscalar components of $j^\mu_\g$.  

 The general form of the  coupling between a  new vector boson $(X)$ and quarks is expressed in terms of a 
different linear combination of $j^\mu_s$ and $j^\mu_v$~\cite{Feng:2016ysn,Feng:2020mbt}: 
\bea
j^\mu_X=\overline{Q} \g^\m\left({\ve_s \over 6}-{\ve_v \t_3\over 2}\right)Q = \ve_s j^\mu_s -\ve_v j^\mu_v,\label{X}
\eea
where $\ve_s$ and $(-\ve_v)$ are the ratios between the $X$ and $\g$ coupling constants in the isoscalar and isovector components. When $\ve_s \approx \ve_v$, $X$ is considered to be protophobic, because the $X$-proton charge-coupling would be much smaller than the $X$-neutron one. In fact, with $\ve_s= \ve_v$ the $X$-$p$ charge coupling vanishes  because there are two $u$  and one $d$ valence quarks in proton, but $X$-$n$ charge coupling is $\ve_v$ times that of $\g$-$p$. 

Comparing \eq{g} and \eq{X} shows that apart from the factor $\ve_s$   the isoscalar ($p+n$) current operators of the $\g$ and $X$ are the same, but (apart from the factor $\ve_v$) the isovector ($p-n$) matrix  current operators differ by a minus sign.

The connection between quark operators and nucleon matrix elements is made explicit using invariance under the  isospin rotation  $P_{cs}\equiv P_y(\pi)$~\cite{Miller:1990iz,Miller:2006tv}, a rotation along $y$-axis by $180^\circ$ in the isospin space, that  interchanges $p$ and $n$ and also (because isospin is an additive quantum number) $d$ and $u$.  Invariance under this rotation gives
\bea 
\langle p | j_s^\mu | p \rangle  =  \langle n| j_s^\mu | n \rangle,\,\quad
\langle p | j_v^\mu | p \rangle  =  -\langle n| j_v^\mu | n \rangle.\label{pcs}\eea
    Hence the nucleon-level  isoscalar $X$-boson current operator is obtained by multiplying the isoscalar photon operator by $\ve_s$ and the nucleon-level isovector $X$-boson current operator is obtained by multiplying the isovector photon operator by $-\ve_v$.

In obtaining \eq{pcs}   isospin symmetry is assumed to be exact. That isospin violation in the nucleon wave function is very small can be anticipated from the small ratio of the neutron-proton mass difference  to their average mass of order $10^{-3}$, and is also verified by explicit calculations, see e.g.~\cite{Miller:1997ya}.)

%

 Therefore, $j^\mu_s$ and $j^\mu_v$'s matrix elements between nucleons are related to the  isoscalar and isovector parts of the EM current's matrix elements (with $u$ as the relevant  Dirac spinor ): 
\begin{align}
& \langle p | j_s^\mu | p \rangle  =  \frac{1}{2}\left( \langle p | j_\g^\mu | p \rangle  + \langle n | j_\g^\mu | n \rangle \right)  \equiv \overline{u}\, \Gamma^\mu_s\, u  \  , \label{gammas} \\ 
& \langle p | j_v^\mu | p \rangle  =   \frac{1}{2}\left( \langle p | j_\g^\mu | p \rangle  - \langle n | j_\g^\mu | n \rangle \right)   \equiv \overline{u}\,  \Gamma^\mu_v \, u \ .  \label{gammav} 
\end{align}
At small values of the momentum transfer the nucleon EM current operators are given by
\begin{align}
& \Gamma^\mu_s  = \frac{\gamma^\mu}{2}  + \l^{(0)}\frac{\sigma^{\mu\nu} i q_\nu }{2 M_N}  \ , \ \Gamma^\mu_v  = \frac{\gamma^\mu}{2}  + \l^{(1)}\frac{\sigma^{\mu\nu} i q_\nu }{2 M_N}   \notag  \\ 
& \mathrm{and} \  J^\mu_\g   =   \overline{N}\left( \Gamma^\mu_s  + \Gamma^\mu_v  \t_3 \right) N  \ , \label{gN}
\end{align}
with $N=(p\ ,n)^T$ as the nucleon field and $J_\g^\mu$ the nucleon-level ($2^\mathrm{nd}$ quantization) current  operator.  With $\lambda^{(0)} = -0.06$ and $\lambda^{(1)} = 1.85$, the magnetic moments $\mu_p = 1 + \lambda^{(0)} + \lambda^{(1)} = 2.79$ and $\mu_n =  \lambda^{(0)} - \lambda^{(1)} = -1.91$.\footnote{It is worth pointing out that the ratio $\mu_n/\mu_p=-0.684 $ is in excellent agreement with the non-relativistic quark-model result of $-2/3$~\cite{Halzen:1984mc}.}

%

Based on Eqs.~(\ref{gammas}),~(\ref{gammav}),~(\ref{pcs}) and~(\ref{X}), the nucleon-level current $J_X^\mu$ can be written as  
\begin{eqnarray}
J^\mu_X & = &  \overline{N}\left(\ve_s \Gamma_s 
 - \ve_v \Gamma_v \t_3 \right) N .  \label{XN}
\end{eqnarray}
This means that while the Dirac ($\g^\m$) coupling of the $X$ to nucleons is protophobic, the Pauli ($\s^{\m\n}$) coupling cannot be so. 
If $\ve_s = \ve_v$, the ratio of neutron to proton $X$-magnetic moments is close to $-3/2$, a value predicted in the non-relativistic quark model. 
%
%

\eq{XN} tells us that, after accounting for kinematic effects (for boson momentum $q^\mu$, $q_\mu q^\mu = M_X^2$ for $X$ and $0$ for $\g$) of the non-zero mass of the $X$ boson, and the different polarization vectors, the isovector (isoscalar) components in the $X$- and $\g$-generating transition matrix elements are related by a simple  factor of $-\ve_v$ ($\ve_s$). As reasoned later, the isovector component dominates over the isoscalar one in all the transitions relevant to this work, so the $X$-production cross section can be inferred from that of the $\g$-production up to an overall factor $\ve_v^2$.   


The next step is to apply the  existing understanding of the EM transitions in $\bee$~\cite{Tilley:2004zz,Zhang:2015ajn,Zhang:2017zap}.  The special feature of the formalism developed for modeling $\lis + p \rightarrow \bee + \g \ (\mathrm{or}\ e^+ + e^-)$ in Ref.~\cite{Zhang:2017zap} is that the effects of non-resonant $\g$ production via an $E1$ electric dipole operator is included along with a magnetic dipole $M1$ induced production that goes through intermediate excited states ($\bee^\ast$). After that, the relation between \eq{gN} and \eq{XN} will be exploited to compute $X$-production cross sections.

\begin{figure}
	\centering
	 \includegraphics[width=0.45 \textwidth]{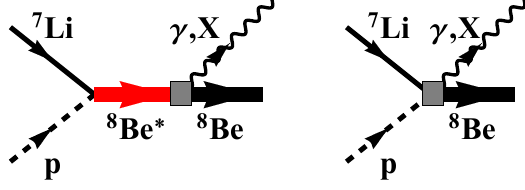}
     \caption{The Feynman diagrams for the $M1$ and $M1^X$ (left), and $E1$ and $E1^X$ (right) transitions. In the left diagram, the intermediate $\bee^\ast$ states are the two $1^+$ resonances.} \label{fig:FeyEM}
\end{figure}

The photon-production matrix element of the $J_\g^\mu$ operator between the initial $p$-$\lis$(${3\over2}^-$) system and the $\bee$ ($0^+$) ground  state is  given by $\langle ^8\mathrm{Be}; -\vec{q} | {J}_\g^{\mu}(\vec{q})| \lis+p;\, a,\,\sigma,\, \vec{p} \rangle$, with $a$ and $\sigma$ as $\lis$ and proton spin projections and  $\vec{q}$ as the (virtual) photon momentum. (From now on, the physical variables in bold fonts, such as $\vec{q}$ denote  3-dimensional vectors.)  This matrix element has various components, labeled by $U_{\lambda S L}$~\cite{Zhang:2017zap}, with $\lambda$, $S$, and $L$ labeling the  $\g$'s multipolarity $(\l)$,  the total spin ($S$) and orbital angular momentum ($L)$ in the initial state. 

The $\g$ production proceeds by either direct  proton capture on $\lis$ (see the right diagram in Fig.~\ref{fig:FeyEM}) or by populating the two intermediate $1^+$ excited states \cite{Tilley:2002vg,Tilley:2004zz} of $\bee$ at relevant beam energies (see the left diagram in Fig.~\ref{fig:FeyEM}).  The properties of the two $1^+$ resonances  are shown  in Fig.~\ref{fig:leveldiag} and Table.~\ref{tab:respara}. Since the scattering energy between $p$ and $\lis$ considered here is very low,  only $s$ and $p$ wave initial states  ($L=0,1$) need to be considered, while its total spin $S= 1, 2$.  Parity and angular momentum conservation leads to selection rules that require only three amplitudes: $\usl{10}$ for $E1$, $\usl{11}$ and $\usl{21}$ for $M1$. The role of the  $E2$ transition is negligible~\cite{Zhang:2017zap} and ignored  here. 

\begin{table}
 \centering
   \begin{tabular}{|c|c|c|c|} \hline
           & $E_{\ms \left(i\right)}$ (MeV) & $\Gamma_{\gamma {\ms \left(i\right)}}$  (eV) & $\Gamma_{\ms \left(i\right)}$ (keV)  \\  \hline
     $i=0$ & 0.895             & $1.9 (\pm 0.4) $              & $ 138(\pm 6) $          \\   \hline
		 $i=1$ & 0.385             & $ 15.0 (\pm 1.8) $            & $10.7 (\pm 0.6)$        \\   \hline
   \end{tabular}
   \caption{From the left to right: the approximate isospin ($i$), excitation energies, EM and strong decay  widths of the $\bee$'s two $1^+$ resonances \cite{Tilley:2004zz}.} \label{tab:respara}
\end{table}

The basic difference  between $X$-boson and $\g$ production is that $M_X$ is non-zero (and is around 17 MeV~\cite{Feng:2016jff}). Therefore, the $X$   
 has three independent polarizations $\tilde{\lambda} = \pm 1, 0$.  We follow Ref.~\cite{Feng:2020mbt} to apply the $q^\mu \varepsilon^{X}_{\mu}(q, {\tilde{\lambda}}) = 0$ constraint~\cite{Weinberg:1995mt}\footnote{Different beyond-standard-model theories have been constructed for massive dark photons~\cite{Feng:2016ysn,Feng:2020mbt}. There, Proca-type lagrangians have been employed (e.g., Eq.~(54) in Ref.~\cite{Feng:2016ysn}) for the new particle, which can be considered as    gauge-fixed versions of the Stueckelberg action\cite{Kors:2004dx}.}  with $\varepsilon^{X}_{\mu}(q,{\tilde{\lambda}})$ as the polarization vector. Based on the derivation of the vector-current multipoles---as a part of the electroweak current---of many-nucleon systems [see Eqs.~(7.20)~(45.12) and~(45.13) in Ref.~\cite{Walecka:1995mi}], we can see that (1) for the transverse polarizations, the corresponding $E1^X$ and $M1^X$ multipoles are defined in the same way as $E1$ and $M1$ multipoles; (2) in addition, there are longitudinal multipoles (e.g., $L0^X$ and $L1^X$) that couples to the $\tilde{\lambda}=0$ polarization, and $J^0_X$-charge-induced multipoles ($C0^X$ and $C1^X$). Since $L$ and $C$ multipoles don't contribute in $\g$ production, no direct information can be drawn for these multipoles from the $\g$ production data. However, in both $X$ and $\g$ productions, their momenta (with $\omega$ as their energy), 
\begin{align}
|\vec{q}| = \begin{cases}
\!  \sqrt{\omega^2 - M_X^2} \sim O(1)\ \text{MeV for $X$} \\ 
\! \omega   \sim O(10) \  \text{MeV for $\g$, }  
\end{cases}  
\end{align}   
are much smaller than the inverse of the nuclear length scale ($\sim O(10^2)$ MeV). In this region, as discussed in the Appendix ~\ref{app:multipoles}, 
$L1^X$ and $C1^X$ are directly related to the $E1^X$. In the following discussion of multipoles, we  focus on  the $E1$, $M1$, $E1^X$ and $M1^X$ at the $|\vec{q}| \to 0$ limit, and  comment on  $C0^X$ and $L0^X$ in the end.

Before going into the reaction formalism~\cite{Zhang:2017zap} which uses $\lis$ and $p$ as fundamental degrees of freedom, it is worth understanding the isospin structure of the $\usl{SL}$ on the nucleon level. It provides the key relationship between $\g$ and $X$ production amplitudes. 

The single-nucleon electric and magnetic multipole operators, derived from the $J^\mu_\g$ current in \eq{gN}, are well-known (e.g., see Eqs.~(5.45) and~(5.71) in Ref.~\cite{Lawson1980} \footnote{Note the convention of nucleon isospin multiplet in Ref.~\cite{Lawson1980} is $N= (n, p)^T$, which is different from ours in Eq.~(\ref{gN}).}). The $E1$  operator is given by
\begin{eqnarray}
\Eone^\g & = & e_\mathrm{EM}\sqrt{\frac{3}{4\pi}}\sum_{i=1}^A \vec{r}_{(i)} \frac{\tau_{(i),3}}{2} \ . \label{e1}
\end{eqnarray}
The summation index  ${(i)}$ labels the nucleons inside the $\bee$ system.  The operator  $\Eone^\g$ is explicitly isovector. The  $M1$ transitions are governed by the operator
\begin{align}
\Mone^\g  =  & \sqrt{\frac{3}{4\pi}} \frac{e_\mathrm{EM}}{2M_N} \sum_i \left[ \left( \l^{(1)}+ \frac{1}{4} \right) \vec{\sigma}_{(i)} \tau_{(i),3} \right.\notag  \\
&  \left.+  \left( \l^{(0)}+ \frac{1}{4} \right) \vec{\sigma}_{(i)} +\frac{1}{2} \vec{J}_{(i)}\left(1+\tau_{(i),3}  \right)  \right] \notag \\ 
        \overset{\mathrm{here}}{\approx} & 
       \sqrt{\frac{3}{4\pi}} \frac{e_\mathrm{EM}}{2M_N} \sum_i \left[ \left(\l^{(1)}+ \frac{1}{4}  \right) \vec{\sigma}_{(i)} + \frac{1}{2} \vec{J}_{(i)} \right] \tau_{(i),3}        \label{m1}   
\end{align}
$\Mone$ is simplified in \eq{m1} based on that (1) the matrix element of the total angular momentum  $\vec{J}=\sum_i \vec{J}_{(i)}$ (assuming the nucleus is made only of nucleons)  between the initial resonances and the final state  are zero, because  $\vec{J}$ does not connect states with different angular momentum; and (2) numerically $|\l^{(1)}+ \frac{1}{4} |\,(=2.10)\,\gg |\l^{(0)}+ \frac{1}{4} |\,(=0.19) $. 
These expressions for $\Eone$ and $\Mone$ are corrected by two-body meson exchange currents that are mainly  transverse and  isovector~\cite{Lawson1980,Pastore:2014oda}. Therefore both $E1$ and $M1$ transitions here are isovector in nature\footnote{The $M1$ transition has been carefully examined in Ref.~\cite{Feng:2016ysn} which also concludes that it is dominated by the isovector component.}.
 
As mentioned above, $E1^X$ and $M1^X$ are defined in the same way as $E1$ and $M1$ but with $J_\g^\mu \to J_X^\mu$~\cite{Walecka:1995mi}.  The resulting $E1^X$ and $M1^X$ transition operators for the $X$ production are obtained  using \eq{XN} (i.e., by  multiplying the isoscalar and isovector components  in both $\Eone$ and $\Mone$ by $\ve_s$ and $-\ve_v$ respectively):  
\begin{eqnarray}
\Eone^X & = & - \ve_v \Eone^\g \label{Xe1} \\ 
\Mone^X & = & \sqrt{\frac{3}{4\pi}} \frac{e_\mathrm{EM}}{2M_N} \sum_i \left[-\ve_v \left( \l^{(1)}+ \frac{1}{4} \right) \vec{\sigma}_{(i)} \tau_{(i),3} \right.
\notag \\
        &&  \left. + \ve_s \left( \l^{(0)}+ \frac{1}{4} \right) \vec{\sigma}_{(i)} +\frac{1}{2} \vec{J}_{(i)}\left(\ve_s -\ve_v \tau_{(i),3}  \right)  \right] \notag \\ 
       & = & \sqrt{\frac{3}{4\pi}} \frac{e_\mathrm{EM}}{2M_N} \sum_i \left\{\ve_s \left( \l^{(0)}+ \frac{1}{4} \right) \vec{\sigma}_{(i)}  \right. \notag \\
        &&   \left. -\ve_v\tau_{(i),3}  \left[ \left(\l^{(1)}+ \frac{1}{4}  \right) \vec{\sigma}_{(i)} + \frac{1}{2}  \vec{J}_{(i)}  \right]   \right\} \label{Xm1_1} \\ 
       & \approx & -\ve_v \Mone^\g     \label{Xm1}       
\end{eqnarray}
The approximation in Eq.~(\ref{Xm1}) would only fail if the isoscalar piece in $\Mone^X$ is greater or comparable than the isovector piece in size, i.e., 
\begin{eqnarray}
 \left\vert  \frac{\ve_s}{\ve_v}\right\vert \gtrsim  \left\vert \frac{\l^{(1)}+ \frac{1}{2}}{\l^{(0)}+ \frac{1}{4}}\right\vert \approx 12. \label{eqn:breakapprox} 
 \end{eqnarray} 
 The ${1}/{2}$ in the $|\l^{(1)}+ {1}/{2}|$ results from combining the spin part of the $\sum_i \tau_{(i),3} \vec{J}_{(i)}$ piece [$= \sum_i \tau_{(i),3} \left(\vec{L}_{(i)} + \vec{\sigma_{(i)}}/2\right)$] with the $ \sum_i \tau_{(i),3} \vec{\sigma}_{(i)}$ piece; the $\vec{L}_{(i)}$ part in the former piece is neglected, because it is either $0$ or $1$ according to shell model and thus its contribution is much smaller than than that of the spin part.
  
Accepting the condition  of \eq{eqn:breakapprox}  would require $X$-proton and -neutron to have almost the same coupling strength, which contradicts $X$ being protophobic. (Note according to Ref.~\cite{Feng:2016ysn}, $\ve_s/\ve_v \lesssim 3 $.)  Moreover, including the two-body current contribution to $\Mone^X$ would further increase~\cite{Pastore:2014oda} the dominance of the isovector component over the isoscalar one, and thus makes Eq.~(\ref{Xm1}) a better approximation.    

\emph{In summary, the $E1^X$ and $M1^X$ operators for  protophobic $X$ boson production are (to an excellent approximation) simply proportional to those for the $\g$ production, with  an overall factor $-\ve_v$.}

\begin{figure}
		\includegraphics[width=.4\textwidth]{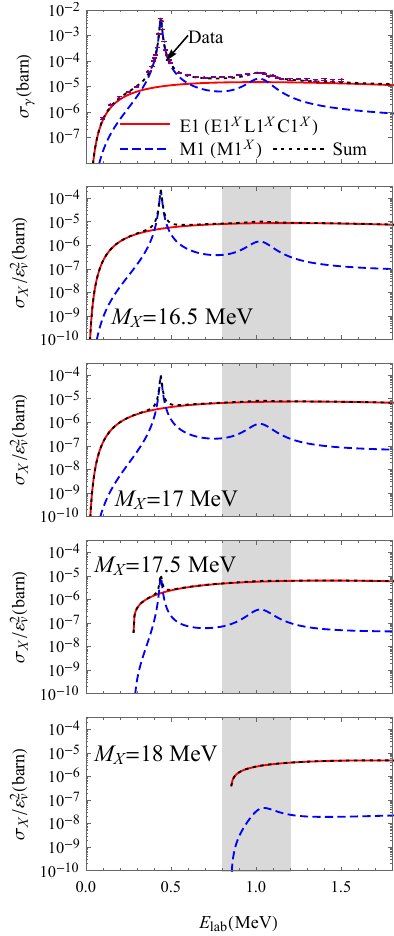} 
     \caption{Top: $\sigma_\g$ {\it vs.} the proton's lab energy $E_\mathrm{lab}$. The data are from Ref.~\cite{Zahnow1995}.  The lower panels: $\sigma_X/\ve_v^2$ for different values of $M_X$. The shaded regions cover the four measured energies~\cite{Krasznahorkay:2015iga}. The legends are shared by all the panels; $E1$ ($E1^X$ $L1^X$ $C1^X$), $M1$ ($M1^X$), and their sum are shown as sold (red), dashed (blue), and dotted (black) lines.} \label{fig:SfacPhotonandX}
\end{figure}

Next we briefly describe our effective field theory  (EFT) inspired model~\cite{Zhang:2017zap} for  $\g$ production, which provides a good description of  the  cross section data~\cite{Zahnow1995}, and the space anisotropy data~\cite{Schlueter1964,Mainsbridge1960}. The model uses  $\lis$ and $p$ as fundamental degrees of freedom to construct the appropriate Lagrangian, so that the model reproduces the properties of nuclear resonances near $\lis$-$p$ threshold, including both MIS and MIV $1^+$ states. Appropriate EM transition vertices are then constructed to describe both direct EM capture process  and the radioactive decay of resonant states populated by $\lis$-$p$ scattering. Their Feynman  diagrams can be found in Fig.~\ref{fig:FeyEM}. The former has smooth dependence on the beam energy while the latter shows resonant behavior. Both components can be qualitatively identified in the $\g$ production data, as shown in the top panel (purple error bars) in Fig.~\ref{fig:SfacPhotonandX}.

The next step is to separate the $E1$ and $M1$ contributions to the $\g$-production cross section and then use the relations in Eqs.~(\ref{Xe1}) and~(\ref{Xm1}) to obtain the $E1^X$ and $M1^X$ contributions to the $X$-boson production.
 One may immediately  expect that the $E1^X$ contributions will be substantial if the $E1$ and $M1$ contributions are comparable.
 This is important because  the observed enhancement of \ep pair-production is associated only   with an   $M1^X$ transition.
  
 The differential cross section can be computed~\cite{Zhang:2017zap} via 
\begin{eqnarray}
\frac{d \sigma_{\ms \gamma, X} }{ d \cos\theta} = \frac{\mr}{ 4 \pi } \frac{q}{p} \frac{1}{8} \sum_{a,\sigma,\tilde{\lambda}} |\mathcal{M}_{\ms \gamma, X}|^2 \ . \label{eqn:Xsecgamma}
\end{eqnarray}
$\mathcal{M}_{\ms \gamma, X}$ is the reaction amplitude depending on polarizations $\tilde{\lambda}$ and nuclear spin projections $a$ and $\sigma$; $M$ the reduced mass between $\lis$ and proton; $\theta$ the angle between boson momentum $\vec{q}$ and beam direction in the CM frame; $q \equiv |\vec{q}|$; $p\equiv \sqrt{2 M E}$ ($E$ as the CM  initial-state kinetic energy  with  $E=7/8 E_\mathrm{lab}$). For both productions, the boson energy $\omega\equiv q^0 = E + E_\mathrm{th}$ ($E_\mathrm{th}$ as the $\lis$-$p$ threshold energy wrt the $\bee$ ground state, see Fig.~\ref{fig:leveldiag}), ignoring the  final state $\bee$'s very small recoiling energy. Note for $\g$, $\omega = q$, while for X, $\omega=\sqrt{M_X^2 + q^2}$.  
  
For $X$ production, $\sum_{\tilde{\lambda}} \varepsilon^X_{\mu}\varepsilon^X_\nu = -\left(g_{\mu\nu} - {q_\mu q_\nu}/{M_X^2}\right)$, since $q^\mu \varepsilon^{X}_{\mu}(q, {\tilde{\lambda}}) = 0$~\cite{Feng:2020mbt}. $\sum_{a,\sigma,\tilde{\lambda}} |\mathcal{M}_{\ms X}|^2$  becomes
\begin{align}
  \sum_{a\sigma\tilde{\lambda}}J_{X\mu} J_{X\nu}^{\ast}\, \varepsilon^{\mu}\varepsilon^\nu 
 = \sum_{a\sigma}J_{X,i} J_{X}^{\ast\, j} \left(\delta_j^i -\frac{q_j q^i}{\omega^2}\right), \label{eqn:Xsec}
\end{align}
with  $J$ now as the currents' matrix elements between nuclear states and $i,j$ as the space indices. The current conservation for which  $J_X^0 = \vec{q}\cdot\vec{J}_X/\omega$ is employed in the derivation. As reasoned above, the pieces in $J_X^\mu$ corresponding to $E1^X$ and $M1^X$ can be derived by multiplying the corresponding pieces in $J_\g^\mu$ by $-\ve_v$.
(The latter's expression in terms of $U_{\l SL}$ can be found in Eq.~(3.1) in Ref.~\cite{Zhang:2017zap}.) 
In addition, Appendix~\ref{app:multipoles} shows that the contributions of $L1^X$ and $C1^X$ associated with $E1^X$ are automatically included in Eq.~(\ref{eqn:Xsec}) as well. 

For an  on-shell photon,  $\omega= q$, so the above formula also applies for $ \sum_{a,\sigma,\tilde{\lambda}} |\mathcal{M}_{\ms \gamma}|^2 $ with $J_X^\mu \to J_\g^\mu$.
 
 The net result, including ($E1$, $M1$) and ($E1^X$, $L1^X$, $C1^X$, $M1^X$) multipoles and evaluating the spin sums, is to arrive at the  following decomposition: 
\begin{eqnarray}&
\sum_{a,\sigma,\tilde{\lambda}} |\mathcal{M}_{\ms \gamma, X}|^2
  = T_0^{\ms \gamma, X} + T_1^{\ms \gamma, X}\, P_1\left(\cos\theta\right) + T_2^{\ms \gamma, X}\, P_2\left(\cos\theta\right), 
\notag\\&\end{eqnarray}
where, $P_{n}$ are the Legendre polynomials, and  
\begin{align}
 T_0^X/\ve_v^2 &=  (3 \omega^2 - {q}^2) |\usl{10}|^2 + \frac{2}{3} {q}^2 \left(\frac{p}{\mr}\right)^2 \bigg[|\usl{11}|^2+|\usl{21}|^2 \bigg]    \ , \label{eqn:T0gammaX} \\ 
 T_1^X/\ve_v^2 &=  2 \sqrt{2} \omega q \left(\frac{p}{\mr}\right)\, \mathrm{Im}\left(\usl{11} \usl{10}^\ast\right)     \ , \label{eqn:T1gammaX} \\
 T_2^X/\ve_v^2 &=  \frac{1}{3} q^2 \left(\frac{p}{\mr}\right)^2 \bigg[|\usl{11}|^2-\frac{1}{5}|\usl{21}|^2 \bigg] \  \label{eqn:T2gammaX} .  
\end{align}
Expressions for $T_n^\g$ (that agree with those in Ref.~\cite{Zhang:2017zap}) are obtained from the above formula  by using $q=\omega$ and setting $\ve_v$ to unity.

Expressions for  $U_{\l SL}$ in terms of EFT coupling parameters are  Eqs.~(3.2), (3.5) and~(3.6) in Ref.~\cite{Zhang:2017zap}. The parameters are fixed by reproducing the photon production data, including total cross section, and ${T_1^\g}/{T^\g_0}$ and ${T_2^\g}/{T^g_0}$ ratios, with  $E_\mathrm{lab}\equiv {8}/{7} E $ below 1.5 MeV. Note the amplitudes $U_{\l SL}$ depend only on $E$, but not $\omega$ or $q$.

\begin{figure}
		\includegraphics[width=.4\textwidth]{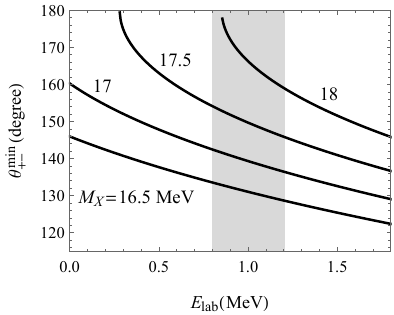} 
     \caption{The minimum values of the pair correlation angles {\it vs.} the proton's lab energy $E_\mathrm{lab}$. Four different cases are plotted with the corresponding masses indicated. The shaded regions again cover the four measured energies~\cite{Krasznahorkay:2015iga}. } \label{fig:thetamin}
\end{figure} 

We now turn to  the results, starting with the $\g$-production cross section shown in the upper panel of Fig.~\ref{fig:SfacPhotonandX}. The model provides  good agreement with the data from Ref.~\cite{Zahnow1995}. For further comparisons between theory and experiment  see Ref.~\cite{Zhang:2017zap}.  The salient features are the two $M$1 resonance contributions, with the lower-energy MIV peak being much higher, and the smooth behavior of those of the  $E1$. Except for the strong peaks at the two $1^+$ resonances, the $E1$ dominates.  

The resonance peaks occur from a two-step process in which the strong interaction connects the initial $\vert\lis,p \rangle$
to the 1$^+$ states which then decay by emitting a photon (see the left diagram in Fig.~\ref{fig:FeyEM}). The relative strengths of the two peaks  naturally arise from \eq{m1}. If the 1$^+$ states were pure isospin eigenstates, the $\Mone$ operator would only connect the  lower-energy state with the ground state. $\g$-production at the higher-energy resonance occurs only because isospin  mixing between the two 1$^+$ states causes the higher-energy state to have an isospin 1 amplitude of $-0.21$~\cite{Pastore:2014oda,Wiringa:2013fia}.  The kinematics together with this ratio can qualitatively explain the ratio of photon widths $\Gamma_{\gamma {\ms \left(i\right)}}$ listed in Table~\ref{tab:respara}~\cite{Feng:2016ysn}.   The difference between the strong decay widths shown in that Table  arises from phase space factors  and the greater importance of the Coulomb barrier at lower energies~\cite{Zhang:2017zap}. The final  $\vert\lis,p \rangle$ states in these strong decays are  equal mixtures of isospin 0 and 1, so the 1$^+$ states' isospin contents do not dictate the strong decay width.

Next turn to the production of $X$ bosons.  Eq.~(\ref{eqn:T0gammaX}) gives the relative magnitude of the total $X$ production cross section i.e., $\sigma_X/\ve_v^2$ and its decomposition into the $ELC1^X$  and $M1^X$ components. (From now on, $ELC1^X$ means the combination of $E1^X$, $L1^X$ and $C1^X$, since they always show up together.) The first term, being proportional to $3\,\omega^2- q^2$, is $ELC1^X$ contributions (for the photon this factor is $2\omega^2$).  The second term gives that of  $M1^X$, whose $q^2$ factor  arises from the  magnetic nature of the interaction. 

The results thus obtained for four different $M_X$ (16.5, 17, 17.5 18 MeV) around the suggested values from Ref.~\cite{Feng:2016jff}  are shown in the lower four panels of Fig.~\ref{fig:SfacPhotonandX}. The shaded regions cover the four different $E_\mathrm{lab}$ values, $0.8, 1.04, 1.1, 1.2$ MeV, that have been measured by the experiment~\cite{Krasznahorkay:2015iga}.  The $X$ can be produced via the dominant $ELC1^X$  component  for almost {\it any}  energy above the kinematic threshold, except around the MIV resonance for $M_X = 16.5$, $17$ and $17.5$ MeV. (The $\lis$-$p$ threshold is $17.26$ MeV above $\bee$'s ground state, and thus no kinetic threshold exists for $M_X =16.5$, $ 17$ and $17.5$ MeV, while such threshold for $M_X=18$ MeV eliminates $X$ production around the MIV resonance.)

One key experimental signal of the $X$ productions is the enhanced \ep detection---from $X$'s decay---in the region of large pair-correlation-angle ($\tpm$) as measured in the lab frame, on the top of the EM-induced pair production background that varies smoothly in the same region~\cite{Krasznahorkay:2015iga, Zhang:2017zap}.  If $M_X \approx \omega$, $\tpm$ is limited to a small window, between $\tpm^\mathrm{min}$ and $180^\circ$, which can be seen based on heavy-particle-decay kinematics (see Ref.~\cite{Feng:2020mbt}). Fig.~\ref{fig:thetamin} plots $\tpm^\mathrm{min}$  against $E_\mathrm{lab}$ for the four different $M_X$ values. For example, with $M_X = 17$ MeV and $0.8 \leq E_\mathrm{lab} \leq 1.2$ MeV (shaded region), the $X$-decay \ep are concentrated in $140^\circ \leq \tpm \leq 180^\circ$; while for other masses, $\tpm$s are in qualitatively similar regions.    

Since the full $X$-production cross section varies smoothly with $E_\mathrm{lab}$ as shown in Fig.~\ref{fig:SfacPhotonandX}, the enhanced \ep  detection in the $\tpm \sim 180^\circ$ region should have been observed across the shaded region. This is in direct conflict with the experimental observation of such enhancement associated only with the higher energy 1$^+$ state, i.e., not seen at $E_\mathrm{lab} = 0.8$ and $1.2$ MeV~\cite{Krasznahorkay:2015iga}.

The dominance of the $ELC1^X$ component  around the MIS resonance and the strong dependence of the $ELC1^X/M1^X$ ratio on the value of $M_X$, as shown in the figure can be understood using  a simple calculation. The ratio  can be inferred from the same ratio for the $\gamma$ production (the phase space factors  canceled  in the ratios). At a given beam energy  $E$,
\begin{eqnarray}
\frac{\sigma_{X, ELC1^X}(E)}{\sigma_{X, M1^X}(E)} =
 \frac{\sigma_{\gamma, E1}(E)}{\sigma_{\gamma, M1}(E)}  \frac{3\omega^2 -q ^2}{2 q^2} \ , \label{rat}
\end{eqnarray}
with $\omega$ as energy for both $X$ and photon, with $q=\sqrt{\omega^2-M_X^2}$.  Now, at the energy of the MIS resonance where the anomaly was observed, Fig.~\ref{fig:SfacPhotonandX} shows $\sigma_{\gamma, E1} \approx 0.7\ \sigma_{\gamma, M1}$. Here $\omega=18.15$ MeV, and  $q=6.36$ MeV for $M_X=17 $ MeV. Then \eq{rat} tells us   that 
\begin{eqnarray}
\left.\frac{\sigma_{X, ELC1^X}}{\sigma_{X, M1^X}}\right\vert_{MIS} = \frac{2 \o^2 + M_X^2}{2(\o^2-M_X^2)} \left.\frac{\sigma_{\gamma, E1}}{\sigma_{\gamma, M1}}\right\vert_{MIS}  \overset{M_X = 17}{\approx} 8.6 \ .  \notag  \\
\end{eqnarray}
The sensitivity to the value of $M_X$ can be seen from the denominator---$M_X$ is close to $\omega$.

The ratio 8.6 is obtained by assuming that \eq{Xm1} is exact.  However to evade this conclusion, $\vert \ve_s / \ve_v \vert$ must  be around or larger than $12$ as shown in \eq{eqn:breakapprox}, which conflicts  with $X$ being protophobic.  

In summary, the results presented in Fig.~\ref{fig:SfacPhotonandX} show that  there would be a  signal of $X$ production due to the $E1^X + L1^X +C1^X$ transitions, i.e., Bremsstrahlung radiation of $X$ boson at all beam energies above threshold.  This  mechanism has a smooth  beam energy $E_\mathrm{lab}$ dependence, while the resonant production diminishes quickly when $E_\mathrm{lab}$ is a few widths away from the resonances.  In fact, given a 17 MeV $X$ boson, the enhancement signal should have been seen at {\it all} four of energies of the ATOMKI experiment~\cite{Krasznahorkay:2015iga}. For a $18$ MeV $X$ boson, although $X$ production around the MIV resonance is eliminated due to kinematic threshold, the smooth Bremsstrahlung component should still be detectable above the MIS resonance. 
However, the experimental observation~\cite{Krasznahorkay:2015iga} of the anomaly is absent below or above the MIS, higher-energy 1$^+$, resonance. 
\emph{Therefore, the explanation of the anomaly in terms of protophobic vector boson $X$ cannot be correct.} 

It is  worth commenting on the $C0^X$ and $L0^X$ multipoles which contribute to the $X$-production but not to the $\g$ production. Although their contributions can not be inferred from the $\gamma$ production, their energy dependences are smooth, because they do not induce transition between $\bee$'s $1^+$ resonance and $0^+$ ground state. Therefore, their contributions \footnote{The $C0^X$ and $L0^X$ cross section should be much smaller than the $E1^X$, because for the former the relative motions in both initial ($^7$Li--$p$) and final states ($^8$Be--$X$) are in p waves, while for the latter both are in s waves. Of course, a definite answer has to be drawn from nuclear microscopic calculations.} enhance the smooth-energy-dependence component in the $X$-production cross section, which further strengthens our basic conclusion. 

Our considerations here are concerned with the $A=8$ system. However, for the $A=4$ system $^4$He, where a signal of $X$-boson production has also been claimed~\cite{Feng:2020mbt,Krasznahorkay:2019lyl}, the Bremsstrahlung terms induced by the ($L0^X$, $C0^X$) multipoles, and the $ELC1^X$ induced resonant productions from the two $1^+$ resonances---about $3.5$ MeV above the experimental measurement~\cite{Krasznahorkay:2019lyl} but with about $6$ MeV widths~\cite{Tilley:1992zz}---will be present. Therefore, given the coupling constants of Ref.~\cite{Feng:2020mbt} one should reasonably expect to have seen a  signal at all beam  energies. Detailed  nuclear calculations of the $L0^X$, $C0^X$, $E1^X$, $L1^X$ and $C1^X$ matrix elements for $A=4$  would be valuable for addressing this issue.

\begin{acknowledgments}
X.Z. was supported in part by the National Science Foundation under Grant No.~PHY--1913069 and by the NUCLEI SciDAC Collaboration under  Department of Energy MSU subcontract RC107839-OSU. G.M. was supported by the US Department of Energy under contract DE-FG02-97ER-41014. G.M. thanks T.~E.~O.~Ericson for useful discussions.
\end{acknowledgments}

\appendix

\section{Multipoles for producing a massive vector boson}\label{app:multipoles}
The reaction amplitude for producing a photon $\g$ or $X$ is
\begin{eqnarray}
\mathcal{M}_{\g,X} =  - e_\mathrm{EM} \ve^{\g,X}_\mu   \langle f | j^\mu_{\g,X}| i \rangle \label{eqn:appM} \ , 
\end{eqnarray} 
with $|i (f)\rangle$ as nuclear states and $j^\mu_{\g,X}$ currents defined in Eqs.~(\ref{g}) and~(\ref{X}).  
Two different ways are employed to describe these amplitudes with $\g$ and $X$'s  $|\vec{q}|\to 0$.

From a low-energy EFT  perspective, the $\g$ and $X$ fields $A_\mu$ and  $X_\mu$ can be treated as external fields~\cite{Gasser:1983yg} in constructing an effective lagrangian. Based on the conservation of $j^\mu_{\g,X}$, the lagrangian density with external fields is invariant under the corresponding \emph{local} symmetry transformation~\cite{Gasser:1983yg}.  The leading order contact terms for producing a $\g$ can be expressed as $\hat{\vec{d}}_\g \cdot (\partial_t \vec{A} -\vec{\partial}A_0)$ for the $E1$, and $\hat{\vec{\mu}}_\g \cdot (\vec{\partial}\times \vec{A})$ for the $M1$. $\hat{\vec{d}}_\g$ and $\hat{\vec{\mu}}_\g$ are electric and magnetic dipole operators, which can be expressed in terms of nuclear cluster fields. For example, in Eqs.~(2.4) and~(2.5) in the Ref.~\cite{Zhang:2017zap}, both operators are constructed using $^7$Li and proton fields. The lagrangian terms for $X$ production take the same forms: $\hat{\vec{d}}_X \cdot (\partial_t \vec{X} -\vec{\partial}X_0)$ and $\hat{\vec{\mu}}_X \cdot (\vec{\partial}\times \vec{X})$, due to the local transformation invariance. The $\hat{\vec{\mu}}_X$  term again corresponds to the $M1^X$, but the $\hat{\vec{d}}_X$ term now has the $E1^X$ $L1^X$ and $C1^X$ contributions, since $X$ is massive. Therefore in Eq.~(\ref{eqn:Xsec}), $ \langle f | j^\mu_{X}| i \rangle$, as derived from  $ \langle f | \hat{\vec{d}}_X| i \rangle$ and  $ \langle f |\hat{\vec{\mu}}_X | i \rangle$, automatically includes the $E1^X$ $L1^X$ and $C1^X$'s contributions together and the $M1^X$'s respectively.

Note the basic argument of this paper is that  $\hat{\vec{d}}_X$  and $\hat{\vec{\mu}}_X$ are proportional to  $\hat{\vec{d}}_\g$ and $\hat{\vec{\mu}}_\g$ respectively because of the dominance of the isovector component. 

On the nucleon level, the electroweak current multipoles have been derived in Ref.~\cite{Walecka:1995mi}. The vector current mulitpoles, including transverse $E^X$ and $M^X$, longitudinal $L^X$, and the charge-induced $C^X$ [Eqs.~(45.12) and~(45.13) in Ref.~\cite{Walecka:1995mi}], can be directly applied here for the $X$ production. By comparing the ($E^X$, $M^X$) to the ($E$ $M$) defined in the Eq.~(7.20) of Ref.~\cite{Walecka:1995mi}, we see that they can be changed into each other by $j_X^\mu \leftrightarrow j_\g^\mu$. Moreover, the $|\vec{q}|\to 0$ limits of these multipoles are in the Eqs.~(45.35)--(45.37) of Ref.~\cite{Walecka:1995mi}. It can be easily checked that in this limit, the relationship between $E1^X$ and ($L1^X$, $C1^X$) are those given by the effective interaction $\hat{\vec{d}}_X \cdot (\partial_t \vec{X} -\vec{\partial}X_0)$ discussed above. I.e., this effective coupling includes all the contributions from the $E1^X$ $L1^X$ and $C1^X$ multipoles.


\end{document}